\documentclass[twocolumn,secnumarabic,amssymb, nobibnotes, aps, prc]{revtex4-2}

\usepackage{bm}
\usepackage{color}
\usepackage{amssymb}
\usepackage{amsmath}
\usepackage{dcolumn}
\usepackage{colortbl}
\usepackage{graphicx}
\usepackage{hyperref}
\usepackage{mathrsfs}
\usepackage{multirow}
\usepackage[table,xcdraw]{xcolor}

\usepackage{url}
\usepackage{ulem}
\usepackage{diagbox}
\usepackage{rotating}
\usepackage{enumitem}

\begin{document}

\title{Density-dependent relativistic mean-field model for $ \Xi^{-} $ hypernuclei}

\author{Shi Yuan Ding}
  \affiliation
  {MOE Frontiers Science Center for Rare Isotopes, Lanzhou University, Lanzhou 730000, China}
  \affiliation
  {School of Nuclear Science and Technology, Lanzhou University, Lanzhou 730000, China}

\author{Ting-Ting Sun}
  \affiliation
  {School of Physics, Zhengzhou University, Zhengzhou 450001, China}

\author{Bao Yuan Sun\footnote{
  Corresponding author (Email: sunby@lzu.edu.cn)}}
  \affiliation
  {MOE Frontiers Science Center for Rare Isotopes, Lanzhou University, Lanzhou 730000, China}
  \affiliation
  {School of Nuclear Science and Technology, Lanzhou University, Lanzhou 730000, China}

\begin{abstract}
In hypernuclear systems, interactions involving nucleons and hyperons are intricately influenced by the surrounding particles, particularly by the density and the isospin feature of the nuclear medium. By studying several observed quantities relevant to hypernuclear bulk and single-particle properties, nuclear in-medium effects and the non-perturbative nature of the strangeness-bearing nuclear force could be revealed. In this work, the relativistic mean-field (RMF) theory is adopted to describe the structure of several typical $\Xi^{-}$ hypernuclei. New sets of $\Xi N$ effective interactions, by taking a density-dependent meson-nucleon/hyperon coupling perspective, are developed by fitting experimental data on the $\Xi^{-}$ hyperon $1s$ and $1p$ state separation energy of $^{15}_{\Xi^{-}}$C as well as the $1p$ state separation energy of $^{13}_{\Xi^{-}}$B. It is found that the density-dependent behavior of meson-hyperon coupling strengths sensitively affects the description of hyperon single-particle levels, which play a crucial role in the consistent description of the theoretical separation energies with experimental data. In fact, the density-dependent meson-baryon coupling strengths introduce additional rearrangement contributions to the hyperon self-energy. Correspondingly, detailed forms of density dependence in these coupling strengths and different considerations of meson-baryon coupling channels will impact the hyperon single-particle properties within hypernuclei. Especially with the additional inclusion of the isovector scalar $ \delta $ meson, the significant enhancement of rearrangement terms in the effective interaction DD-ME$\delta$ impacts the shape of the hyperon potential and alters the characteristics of the isovector channel dynamics balance in the effective nuclear force. As the difference in the $\Xi^{-}1s$ separation energy of $^{15}_{\Xi^{-}}$C remains large among three sets of $\Xi N$ effective interactions, a possible explanation to understand the experimental results is taken further by considering the mixing between the $\Xi^{-}$ state in $^{14}$N and the $\Xi^{0}$ state in $^{14}$C. Relevant research underscores the importance of precisely accounting for in-medium effects in hyperon-nucleon interactions and incorporating a more comprehensive set of meson-exchange degrees of freedom in effective nuclear forces, offering a potential solution for more self-consistently describing the featured hyperon single-particle behavior of various hypernuclei and for reducing uncertainties in theoretical descriptions.
\end{abstract}


\maketitle

\section{Introduction}
Hypernuclear physics presents a new direction in exploring the nuclear chart and provides a unique tool for extending our present knowledge of conventional nuclear physics into the SU(3)-flavor sector \cite{Lenske2018PPNP98.119}. With strangeness degrees of freedom, hyperons are free from the constraints of the Pauli exclusion principle among nucleons, analogous to impurities moving deep into the nucleus. This affords a unique perspective for the study of baryon-baryon interactions in nuclear mediums \cite{Yamamoto1994Prog.Theor.Phys.Suppl.117.361, Gibson1995Phys.Rep.257.349, Epelbaum2009Rev.Mod.Phys.81.1773, Hiyama2009Prog.Part.Nucl.Phys.63.339}, which is essential for the understanding of nuclear structures as well as neutron-star matter as hyperons would emerge at high densities. In the past decades, considerable experimental data have been garnered for single-$ \Lambda $ hypernuclei with strangeness $S=-1$ from light to heavy mass ranges \cite{Hashimoto2006Prog.Part.Nucl.Phys.57.564, Feliciello2015Rep.Prog.Phys.78.096301, Gal2016Rev.Mod.Phys.88.035004}. For hypernuclear systems with multi-strangeness $S=-2$, i.e., $\Lambda\Lambda$ or $\Xi$ hypernuclei, due to the smaller production cross section as the shorter lifetime of a $\Xi$ hyperon, experimental information has only gradually started to be obtained in recent years and predominantly in the light mass range \cite{Feliciello2015Rep.Prog.Phys.78.096301, Gal2016Rev.Mod.Phys.88.035004, Nakazawa2015Prog.Theor.Exp.Phys.2015.033D02, Hayakawa2021PRL126.062501, Yoshimoto2021PTEP2021.073D02}.

Experimentally, to produce doublestrangeness $S=-2$ hypernuclei, the $(K^{-}, K^{+})$ reaction is an effective method, which transfers two strangeness and charge units to the target nucleus \cite{Feliciello2015Rep.Prog.Phys.78.096301, Gal2016Rev.Mod.Phys.88.035004}. In recent years, with the advancement of radioactive ion beam facilities and experimental analysis techniques such as the ``emulsion-counter hybrid method" and the ``overall scanning method" \cite{Yoshida2017Nucl.Instrum.Methods-Phys.Res.A847.86}, a few $ \Xi^{-} $ hypernuclei events have been detected. Regarding the $^{12}_{\Xi^{-}}$Be and $^{13}_{\Xi^{-}}$B hypernuclei, several early-stage emulsion data are available \cite{Dover1983Ann.Phys.146.309, Aoki1993Prog.Theor.Phys.89.493, Aoki1995PLB355.45,  Fukuda1998PRC58.1306, Khaustov2000PRC61.054603, Yamaguchi2001Prog.Theor.Phys.105.627}. An empirical value of $B_{\Xi^{-}}=4.5$ MeV has been adopted for $^{12}_{\Xi^{-}}$Be ($^{11} $B+$ \Xi^{-}$), which was derived from few-body calculations assuming a Woods-Saxon potential with depth $V=-14$ MeV for the $ \Xi N $ interaction \cite{Khaustov2000PRC61.054603, Hiyama2008PRC78.054316}. Recently, in the J-PARC-E05 experiment, a preliminary single-$ \Xi^{-} $ separation energy $B_{\Xi^{-}} $ in $^{12}_{\Xi^{-}}$Be was reported to be $ 6.3 $ MeV in the one-peak interpretation while $ 9 $ MeV and $ 2 $ MeV in a two-peak interpretation \cite{Nagae2019AIP.Conf.Proc.2130.020015}. For $^{13}_{\Xi^{-}}$B ($^{12} $C+$ \Xi^{-}$), two possible values of the single-$ \Xi^{-} $ separation energy, i.e., $B_{\Xi^{-}} = 0.82\pm0.17$ MeV and $B_{\Xi^{-}} = 0.82\pm0.14$ MeV, have been reported by the KEK E176 collaboration team \cite{Aoki2009NPA828.191, Nakazawa2015Prog.Theor.Exp.Phys.2015.033D02}, which corresponds to $ \Xi^{-}{1p} $ nuclear states that evolve from $ 2P $ atomic states upon adding a strong-interaction $ \Xi^{-} $ nuclear potential \cite{Friedman2021PLB820.136555}. However, due to limitations in experimental techniques and analysis methods, significant uncertainty persists in the $ \Xi N $ interactions as well as in the single-$ \Xi^{-} $ separation energies.

In 2015, the famous KISO event related to the reaction $ ^{14} $N+$ \Xi^{-}$\textrightarrow $ ^{15}_{\Xi^{-}} $C\textrightarrow $ ^{10}_{\Lambda} $Be$ +^{5}_{\Lambda} $He was observed in the KEK E373 emulsion experiment, which provided a direct evidence for a deeply bound $ \Xi^{-} $ hypernuclear system and attractive $ \Xi N $ interactions for the first time \cite{Nakazawa2015Prog.Theor.Exp.Phys.2015.033D02}. Two possible single-$ \Xi^{-} $ separation energies, i.e., $B_{\Xi^{-}} = 4.38\pm0.25 $ MeV and $B_{\Xi^{-}} = 1.11\pm0.25 $ MeV, were proposed, corresponding to the cases with $ ^{10}_{\Lambda} $Be in the ground and first excited states. Subsequently, the binding energy for $ ^{10}_{\Lambda} $Be was further revised \cite{Gogami2016PRC93.034314}, and the $ B_{\Xi^{-}} $ of $ ^{15}_{\Xi^{-}} $C in the KISO event was updated to $ 3.87\pm0.21 $ MeV and $ 1.03\pm0.18 $ MeV, respectively \cite{Hiyama2018Ann.Rev.Nucl.Part.Sci.68.131}. Recently, two new events, named KINKA and IRRAWADDY, were identified in the KEK E373 and J-PARC E7 experiments, which determined the single-$ \Xi^{-} $ separation energy to be $ 8.00 \pm 0.77 $ MeV or $ 4.96 \pm 0.77 $ MeV corresponding to $ 1s $, $ 1p $ state in KINKA event while $ B_{\Xi^{-}} = 6.27 \pm 0.27 $ MeV corresponding to $ 1s $ state in the IRRAWADDY event \cite{Yoshimoto2021PTEP2021.073D02}.

Although the amount of current experimental data has increased, significant uncertainty in the $\Xi N$ interaction remains due to the limited precision of these data. Various theoretical models have been developed to investigate $ \Xi^{-} $ hypernuclear structures, such as the chiral effective field theory \cite{Hoai2021EPJA57.339}, the optical potential methodology \cite{Friedman2021PLB820.136555, Friedman2022EPJWebConf.271.03002, Friedman2023PLB837.137640}, the Gaussian expansion method \cite{Hiyama2008PRC78.054316, Hiyama2020PRL124.092501}, the antisymmetrized molecular dynamics model \cite{Isaka2024PRC109.044317}, the Skyrme-Hartree-Fock (SHF) theory \cite{Jin2020EPJA56.135, Guo2021PRC104.L061307}, the quark-meson coupling model \cite{Tsushima1998NPA630.691, Guichon2008NPA814.66, Shyam2019arXiv1901.06090}, the quark mean-field model \cite{Hu2017PRC96.054304, Hu2022JPG49.025104}, and the relativistic mean-field (RMF) model \cite{Jennings1994PRC49.2472, Sun2016PRC94.064319, Liu2018PRC98.024316, Tanimura2022PRC105.044324}. These studies have extensively explored the properties of light $ \Xi $ hypernuclei, including aspects such as the existence of the lightest $ \Xi^{-} $ hypernuclear system \cite{Hiyama2020PRL124.092501}, the decay modes \cite{Sun2016PRC94.064319, Hu2017PRC96.054304, Guo2021PRC104.L061307, Tanimura2022PRC105.044324}, and the effects of deformation \cite{Jin2020EPJA56.135, Guo2021PRC104.L061307}. Due to its capacity to offer a self-consistent and unified description, RMF theory has achieved great success in the description of finite nuclei all across the nuclear chart and nuclear matter \cite{Reinhard1989Rep.Prog.Phys.52.439, Ring1996Prog.Part.Nucl.Phys.37.193, Vretenar2005Phys.Rep.409.101, Meng2006Prog.Part.Nucl.Phys.57.470}. Moreover, it has been extended to encompass the description of hypernuclear systems with strange degrees of freedom \cite{Jennings1994PRC49.2472, Lu2014PRC89.044307, Sun2016PRC94.064319, Liu2018PRC98.024316, Rong2021PRC104.054321, Chen2021Sci.ChinaPhys.Mech.Astron.64.282011, Tanimura2022PRC105.044324, Ding2023CPC47.124103, Yang2024PRC110.054320}. Based on RMF and SHF models, theoretical studies have indicated that $ ^{15}_{\Xi^{-}} $C from the KISO event is in an excited state with the single-$ \Xi^{-} $ hyperon occupying the $ 1p $ orbital \cite{Sun2016PRC94.064319}. This prediction was further supported by the IBUKI event, in which the single-$ \Xi^{-} $ separation energy was observed as $B_{\Xi^{-}} = 1.27\pm0.21 $ MeV \cite{Hayakawa2021PRL126.062501}. These theoretical works provide significant information for the effective $\Xi N$ interaction.

Since the hyperon inside hypernuclei is located in a nuclear medium, the $YN$ interaction is then influenced by the in-medium effects remarkably. Therefore, it deserves to check carefully the influence of different treatments for in-medium effects on the bulk and single-particle properties of $ \Xi^{-} $ hypernuclei. Inspired by microscopic calculations within the Dirac Brueckner-Hartree-Fock theory \cite{Brockmann1992PRL68.3408}, the nuclear in-medium effects are important which can be considered by introducing the density-dependent meson-nucleon coupling strengths, the validity and importance of which have been demonstrated in numerous early studies on finite nuclei and nuclear matter \cite{Typer1999NPA656.331, Hofmann2001PRC64.025804, Hofmann2001PRC64.034314, Niksic2002PRC66.024306, Niksic2002PRC66.064302, Tryggestad2003PRC67.064309}. The resulting density-dependent relativistic mean-field (DDRMF) and the density-dependent relativistic Hartree-Fock (DDRHF) theories incorporates these variable coupling strengths, making the effective nuclear force dependent on the density of the nuclear medium. Consequently, this approach has profound implications for the description of finite nuclear structures from the core to the surface, as well as for the properties of nuclear matter across a range of densities from low to high, and has led to a multitude of significant and intriguing discoveries, such as nuclear symmetry energy \cite{Sun2008PRC78.065805, Long2012PRC85.025806, Zhao2015JPG42.095101, Liu2018PRC97.025801}, nucleon effective masses \cite{Long2006PLB640.150}, liquid-gas phase transition \cite{Zhang2013PLB729.148, Yang2019PRC100.054314, Yang2021PRC103.014304}, equation of state (EOS) of dense matter \cite{Shen2010PRC82.015806}, neutron star \cite{Avancini2007PRC75.055805, Wang2014PRC90.055801}, shell evolution \cite{Wang2013PRC87.047301, Li2016PLB753.97, Liu2020PLB806.135524}, neutron skin effects \cite{Avancini2007PRC75.055805, Avancini2007PRC76.064318}, nuclear mass \cite{Afanasjev2013PLB726.680, Taninah2024PRC109.024321}, and nucleon drip lines \cite{Niksic2002PRC66.024306, Vretenar2005Phys.Rep.409.101, Chen2012PRC85.067301, Afanasjev2013PLB726.680, Afanasjev2016PRC93.054310}. Additionally, the density-dependent couplings change essentially the in-medium equilibrium between attraction and repulsion of nuclear force, which affects the description of nuclear properties at various mass and isospin numbers \cite{Geng2019PRC100.051301, Ding2023CPC47.124103}. For instance, by taking a unique density-dependent form, a new density-dependent effective interaction DD-LZ1 \cite{Wei2020CPC44.074107} has been developed, which solves the common problem of the $ Z = 58,~92 $ pseudo-shell closures in the framework of RMF theory for the first time and shows great advantages in the descriptions of the neutron star crust physics \cite{Xia2022PRC105.045803}. Recently, DDRMF/DDRHF theories have been further applied to study the single-$ \Lambda $ hypernuclei, where the impact of in-medium effects on the hyperon spin-orbit splittings has been discussed \cite{Rong2021PRC104.054321, Ding2022PRC106.054311, Ding2023CPC47.124103}. Therefore, it is essential to further investigate the possible effects of different treatments of the effective nuclear force in the medium on the description of the hypernuclear structure.

Apart from the different treatments of in-medium effects, varying considerations of the meson-baryon couplings also impact the description of the bulk and single-particle properties of hypernuclei. Since the $\Xi$ hyperon is an isovector particle, compared to the $\Lambda$ hyperon, it requires accounting for the additional contributions from its coupling with isovector mesons. In past decades, the isovector-vector $\rho$ meson has been considered in the $\Xi N$ interactions within the RMF models. The results have demonstrated that the $\rho$ meson exerts a significant impact on the single-particle energies, separation energies, and hyperon potentials, particularly in describing hypernuclear systems with $N\neq Z$ \cite{Jennings1994PRC49.2472, Sun2016PRC94.064319, Hu2017PRC96.054304}. Additionally, the role of the $\rho$ meson in single-$\Xi$ hypernuclei should be carefully considered, especially within pure isospin-zero cores where the $\rho$ meson is introduced solely by the $\Xi$ hyperon. However, since only one $\Xi$ hyperon exists, the influence of the $\rho$ meson is spurious in the Hartree approximation and should be removed \cite{Jennings1994PRC49.2472}. Besides, the isovector scalar meson, namely the $\delta$, affects nuclear isospin properties as well, such as the splitting of nucleon's Dirac mass \cite{Liu2002PRC65.045201}. As a fundamental aspect of nuclear forces, the contribution of the isovector scalar $\delta$ meson has been acknowledged in numerous studies for its critical role in comprehending nuclear matter properties and the structure of finite nuclei \cite{Kubis1997PLB399.191, Shao2010PRC82.055801, Huang2015CPC39.105102, Dutra2016PRC93.025806, Qian2018Sci.China-Phys.Mech.Astron.61.082011, Li2022APJ929.183}. For example, when considering the isovector scalar channel, the equation of state (EOS) for neutron star undergoes a certain softening, consequently reducing the maximum mass and radius \cite{Qian2018Sci.China-Phys.Mech.Astron.61.082011}. In calculating the direct Urca processes in neutron stars with hyperons, the $ \delta $ meson leads to a significant enhancement in the total neutrino emissivity, thereby accelerating the cooling rate of neutron stars \cite{Huang2015CPC39.105102}. With further consideration of the coupling effects between the $ \sigma $ and $ \delta $ mesons, a unified framework at the mean-field level to concurrently describe finite nuclei, flow data in the heavy-ion collision, and constraints on the mass-radius relation of neutron star \cite{Li2022APJ929.183}. Therefore, as an indispensable component of the $ \Xi N $ interaction, the impact of the $ \delta $ meson on the $ \Xi $ hypernuclear properties warrants further exploration.

The existing hypernuclear experimental data, while enriching our understanding of hypernuclear structure and baryon-baryon interactions, also pose new challenges for the development of a self-consistent theoretical description. Many models struggle to provide a reasonable description of the diverse experimental results. Therefore, in this work, we will extend the density-dependent relativistic mean-field model, which has already been successfully applied to the description of finite nuclei and nuclear matter properties, to explore the structure of $\Xi$ hypernuclei, with the aim of offering a reasonable description of the experimental results. The essential role of the nuclear in-medium effects and the isovector scalar $ \delta $ meson will be discussed. In Sec. {\ref{Theoretical Framework}}, the theoretical framework is presented. In Sec. {\ref{Results and Discussion}}, the nuclear in-medium effects and the impact of the isovector scalar $ \delta $ meson on the bulk and single-particle properties of hypernuclei will be studied. Finally, a summary will be given in Sec. {\ref{Summary and Outlook}}.

\section{Theoretical Framework}\label{Theoretical Framework}
The formalism of the DDRMF theory with the $ \Xi $ hyperon degree of freedom will be briefly introduce, which starts from the following Lagrangian density
\begin{align}\label{eq:Lagrangian}
 \mathscr{L} = &\sum_{B=N,\Xi}\bar{\psi}_{B}\left(i\gamma^{\mu}\partial_{\mu}-M_{B}-g_{\sigma B}\sigma-g_{\omega B}\gamma^{\mu}\omega_{\mu}\right. \notag\\
 &\left.-g_{\delta B}\vec{\tau}_{B}\cdot\vec{\delta} -g_{\rho B}\gamma^{\mu}\vec{\tau}_{B}\cdot\vec{\rho}_{\mu} \right)\psi_{B}\notag\\
&+\frac{1}{2}\partial^{\mu}\sigma\partial_{\mu}\sigma-\frac{1}{2}m_{\sigma}^{2}\sigma^{2}-\frac{1}{4}\Omega^{\mu\nu}\Omega_{\mu\nu}+\frac{1}{2}m_{\omega}^2\omega^{\mu}\omega_{\mu}\notag\\
&+\frac{1}{2}\partial^{\mu}\vec{\delta}\cdot\partial_{\mu}\vec{\delta}-\frac{1}{2}m_{\delta}^{2}\vec{\delta}^{2}\notag\\
&-\frac{1}{4}\vec{R}^{\mu\nu}\cdot\vec{R}_{\mu\nu}+\frac{1}{2}m_{\rho}^2\vec{\rho}^{\mu}\cdot\vec{\rho}_{\mu}-\frac{1}{4}F^{\mu\nu}F_{\mu\nu}\notag\\
&-\bar{\psi}_{N}e\gamma^{\mu}\frac{1-\tau_{3,N}}{2}A_{\mu}\psi_{N}\notag\\ &-\bar{\psi}_{\Xi}\left(-e\gamma^{\mu}\frac{1+\tau_{3,\Xi}}{2}A_{\mu}+\frac{f_{\omega\Xi}}{2M_{\Xi}}\sigma^{\mu\nu}\partial_{\nu}\omega_{\mu}\right)\psi_{\Xi},
\end{align}
where $M_{B}$ is the baryon mass, $m_{\phi}$ denotes the masses for the $\phi=\sigma,~\omega_{\mu},~\vec{\delta},~\vec{\rho}_{\mu}$ mesons, and $\Omega^{\mu\nu}$, $\vec{R}^{\mu\nu}$ and $F^{\mu\nu}$ are the field tensors of vector mesons $\omega_{\mu}$, $ \vec{\rho}_{\mu}$ and photon $A_{\mu}$, respectively. $\vec{\tau}_{B}$ is the isospin operator with the third component $\tau_{3,N}=1$ for the neutron, $\tau_{3,N}=-1$ for the proton, $\tau_{3,\Xi}=1$ for the $ \Xi^{-} $ hyperon, and $\tau_{3,\Xi}=-1$ for the $ \Xi^{0} $ hyperon. $g_{\phi B}$ represent the meson-baryon coupling strengths, while $ \frac{f_{\omega\Xi}}{2M_{\Xi}} $ denotes the tensor coupling between hyperons and the $ \omega $ field.

In the density-dependent RMF approach, the coupling strengths are determined by baryon-density-dependent functions to phenomenologically introduce the nuclear in-medium effects \cite{Long2006PLB640.150}. Specifically, the coupling strengths between baryons and isoscalar mesons ($\sigma$ and $\omega_{\mu}$) in density-dependent effective interactions adopted in this work are expressed as follows:
\begin{align}\label{eq:coupling_constants1}
g_{\phi B}\left(\rho_{b}\right)=g_{\phi B}(0) a_{\phi B}\frac{1+b_{\phi B}(\xi+d_{\phi B})^2}{1+c_{\phi B}(\xi+e_{\phi B})^2},
\end{align}
where $\xi=\rho_{b}/\rho_{0}$, with $\rho_{0}$ being the saturation density of nuclear matter. The density dependence in DD-ME$ \delta $ \cite{Roca-Maza2011PRC84.054309} for the coupling strengths between baryons and isovector mesons ($ \vec{\rho}_{\mu} $ and $ \vec{\delta} $) is given by Eq. \eqref{eq:coupling_constants1}, whereas in other effective interactions, it is described by
\begin{align}\label{eq:coupling_constants2}
g_{\phi B}\left(\rho_{b}\right)=g_{\phi B}(0) e^{-a_{\phi B} \xi}.
\end{align}
In the above expression, $g_{\phi B}(0)$ corresponds to the free coupling strength at $\rho_b=0$.

In systems exhibiting time-reversal symmetry, the space-like components of the vector fields vanish. Additionally, it is reasonable to presume that nucleon and $\Xi$ hyperon single-particle states are unaffected by isospin mixing, indicating that these states are eigenstates of $\tau_{3,B}$, so that only the third component of $\vec{\rho}_{\mu}$ and $\vec{\delta}$ survives. For convenience, in the following, we shall use $\sigma$, $\omega$, $\rho$, $\delta$ and $A$ to denote the various meson/photon fields.

With the mean-field and no-sea approximations, we can derive the single-particle Dirac equations for baryons, the Klein-Gordon equations for mesons, and the Poisson equations for photon by the variation principle. In the following, the description of $\Xi$ hypernuclei is restricted to the spherical symmetry. Correspondingly, the complete set of good quantum numbers contains the principle one $n$, the total angular momentum $j$ and its projection $m$, as well as the parity $\pi=(-1)^l$ ($l$ is the orbital angular momentum). By taking the quantum number $\kappa$ to denote the angular momentum $j$ and the parity $\pi$, i.e., $\kappa=\pm(j+1/2)$ and $\pi=(-1)^\kappa\text{sign}(\kappa)$, the Dirac spinor $f_{i}(\bm{x})$ of the nucleon or hyperon has the following form with spherical coordinate ($r,\vartheta,\varphi$):
\begin{align}
  f_{n\kappa m}(\bm{x}) =  \frac{1}{r} \left(\begin{array}{c}iG_a(r)\Omega_{\kappa m}(\vartheta,\varphi)\\ F_a(r)\Omega_{-\kappa m}(\vartheta,\varphi) \end{array}\right),
\end{align}
where the index $a$ consists of the set of quantum numbers $(n\kappa) = (njl)$, and $\Omega_{\kappa m}$ is the spherical spinor. Then, the Dirac equations for the nucleons and the $\Xi$ hyperon can be expressed as
\begin{align}
\begin{pmatrix}
\Sigma_{+}^{B}-\varepsilon_{a,B} & \displaystyle-\frac{d}{dr}+\frac{\kappa_{a,B}}{r}+\Sigma_{T}^{B} \\
\displaystyle\frac{d}{dr}+\frac{\kappa_{a,B}}{r}+\Sigma_{T}^{B} & -2M_{B}+\Sigma_{-}^{B}-\varepsilon_{a,B}
\end{pmatrix}
\begin{pmatrix}
G_{a,B} \\ F_{a,B}
\end{pmatrix}=0,
\label{eq:DiracEquation}
\end{align}
and the Klein-Gordon equations for mesons and the Poisson equation for photon read

\begin{align}
  (-\nabla^{2}+m_{\sigma}^{2})\sigma&=-g_{\sigma N}\rho_{s,N}-g_{\sigma \Xi}\rho_{s,\Xi},\label{eq:field equation for sigma}\\
  (-\nabla^{2}+m_{\omega}^{2})\omega&=+g_{\omega N}\rho_{b,N}+g_{\omega \Xi}\rho_{b,\Xi}+\frac{f_{\omega\Xi}}{2M_{\Xi}}\partial_{i}j^{0i}_{T\Xi},\label{eq:field equation for omega}\\
  (-\nabla^{2}+m_{\delta}^{2})\delta&=-g_{\delta N}\rho_{s,N}\tau_{3,N}-g_{\delta \Xi}\rho_{s,\Xi}\tau_{3,\Xi},\label{eq:field equation for delta}\\
  (-\nabla^{2}+m_{\rho}^{2})\rho&=+g_{\rho N}\rho_{b,N}\tau_{3,N}+g_{\rho \Xi}\rho_{b,\Xi}\tau_{3,\Xi},\label{eq:field equation for rho}\\
  -\nabla^{2}A&=+e\rho_{b,N}Q_{N}+e\rho_{b,\Xi}Q_{\Xi}.\label{eq:field equation for photon}
\end{align}
Here, $ \rho_{s,B} $, $ \rho_{b,B} $ and $j^{0i}_{T\Xi}$ represent the scalar, baryon and tensor densities, respectively, the total baryon density is $ \rho_{b}=\rho_{b,N}+\rho_{b,\Xi} $ \cite{Ding2022PRC106.054311, Sun2016PRC94.064319}. In Eq. \eqref{eq:field equation for photon}, $Q_{N}$ and $Q_{\Xi}$ represent $\frac{1-\tau_{3,N}}{2}$ and $-\frac{1+\tau_{3,\Xi}}{2}$, respectively.

The local self-energies in Eq. \eqref{eq:DiracEquation}, denoted as $\Sigma_{\pm}^{B} = \Sigma_{0,B} \pm \Sigma_{S,B}$, comprise vector and scalar terms. Additionally, $\Sigma_{T}^{B}$ incorporates contributions from the tensor component. Notably, $\Sigma_{T}^{B}$ is zero for nucleons, but for hyperons, it specifically originates from the $\omega$ tensor within the hyperon channel \cite{Jennings1994PRC49.2472, Sun2016PRC94.064319}. The scalar self-energy and the time component of the vector self-energy can be expressed as
\begin{subequations}
\begin{align}
  \Sigma_{S,B} &= g_{\sigma B}\sigma+g_{\delta B}\tau_{3,B}\delta,\\
  \Sigma_{0,B} &= g_{\omega B}\omega+g_{\rho B}\tau_{3,B}\rho+eQ_{B}A+\Sigma_{R},
\end{align}
\end{subequations}
In addition, $\Sigma_R$ is the rearrangement term due to the density dependence of the coupling constant, can be expanded as follows:
\begin{align}
  \Sigma_{R}&=\sum_{B}\left(\dfrac{\partial g_{\sigma B}}{\partial \rho_{b}}\rho_{s,B}\sigma + \dfrac{\partial g_{\omega B}}{\partial \rho_{b}}\rho_{b,B}\omega\right.\notag\\
  &+ \left.\dfrac{\partial g_{\delta B}}{\partial \rho_{b}}\rho_{s,B}\tau_{3,B}\delta+\dfrac{\partial g_{\rho B}}{\partial \rho_{b}}\rho_{b,B}\tau_{3,B}\rho\right).
  \end{align}

\section{Results and Discussion}\label{Results and Discussion}
Now we can apply the RMF theory to investigate the bulk and single-particle properties of the $ \Xi $ hypernuclei. To explore the impact of nuclear in-medium effects on the description of hypernuclear structure, several density-dependent RMF models were selected for nucleon-nucleon ($ NN $) interactions, including TW99 \cite{Typer1999NPA656.331}, PKDD \cite{Long2004PRC69.034319}, DD-ME2 \cite{Matsumiya2011PRC83.024312}, DD-MEX \cite{Taninah2020PLB800.135065, Rather2021PRC103.055814}, DD-ME$\delta$ \cite{Roca-Maza2011PRC84.054309}, and DD-LZ1 \cite{Wei2020CPC44.074107}. Among these, the effective interaction DD-ME$\delta$ introduces an additional isovector scalar coupling channel, enabling an effective exploration of its impact on the description of hypernuclear structure. Additionally, the nonlinear RMF effective interaction PK1 \cite{Long2004PRC69.034319} was used for comparison. The Dirac equation is solved in a radial box size of $ R=20 $ fm with a step of $ 0.1 $ fm. For open-shell nuclei, the pairing correlation is addressed using the BCS method. Additionally, the blocking effect is considered for the last valence nucleon or hyperon \cite{Perez-Martin2008PRC78.014304}. For each hypernucleus, we verify the binding energy values by applying the blocking procedure to different nucleon/hyperon orbitals near its Fermi surface, and we select the configuration with the lowest binding energy as its ground state.

\subsection{$\Xi N$ effective interaction in RMF models}
Within the framework of RMF theory, the $ \Xi N $ interaction relates to the coupling strengths among the mesons and $ \Xi $ hyperons involved in the interaction. Specifically, the ratio of isoscalar vector coupling strength $ g_{\omega\Xi}/g_{\omega N} $ is set at $ 0.333 $ based on the n\"{a}ive quark model \cite{Dover1984Prog.Part.Nucl.Phys.12.171}. The isovector vector coupling strength $ g_{\rho\Xi}=g_{\rho N} $ is determined using SU(3) Clebsch-Gordan coefficients \cite{Jennings1994PRC49.2472}. For the ratio of isovector scalar coupling strength $ g_{\delta\Xi}/g_{\delta N} $, a fixed value of 1.000 is employed \cite{Shao2010PRC82.055801}. Furthermore, in accordance with Refs. \cite{Jennings1994PRC49.2472, Sun2016PRC94.064319}, the tensor coupling is considered in the hyperon channel, with a coupling strength of $ f_{\omega\Xi}=-0.4g_{\omega\Xi} $. The ratio of the isoscalar scalar coupling strength $ g_{\sigma\Xi}/g_{\sigma N} $ can be determined by reproducing the experimental data on the separation energies $ B_{\Xi^{-}} $ of the $ \Xi^{-} $ hyperon. Here, $ B_{\Xi^{-}} $ is defined as:
\begin{align}
B_{\Xi^{-}}[A]&\equiv E[n,p,-]-E[n,p,\Xi^{-}]\notag\\
&=E[^{A-1}(Z+1)]-E[^{A}_{\Xi^{-}}Z],
\end{align}
where $E$ gives the binding energy of a hypernucleus or its nucleonic core. For hypernuclei, $ A=Z+N+1 $. Note that in DDRMF models, while the coupling strength between mesons and hyperon/nucleons evolves gradually with baryon density, the values of $g_{\phi \Xi}/g_{\phi N}$ are fixed. Moreover, the mass of the $\Xi^{-}$ hyperon is taken to be $M_{\Xi^{-}}=1321.7$ MeV.

Given the considerable uncertainty in current experimental data on the separation energy of $\Xi^{-}$ hyperons, selecting appropriate fitting targets is crucial for constructing the $\Xi N$ interaction and reliably describing the hypernuclear structure. As emphasized in the introduction, the deeply bound $^{15}_{\Xi^{-}}$C hypernucleus, which was first conclusively discovered in experiments with an attractive $\Xi N$ interaction, is an ideal candidate for determining the $\Xi N$ interaction. The KISO and IBUKI experiments have consistently provided results for the separation energy of $ \Xi^{-} $ hyperons in the $1p$ state of $^{15}_{\Xi^{-}}$C hypernucleus. Thus, the weighted average separation energy $B_{\Xi^{-}}=1.13\pm0.14$ MeV obtained from these two experiments serves as a critical objective for constructing the $\Xi N$ interaction \cite{Hayakawa2021PRL126.062501}. Regarding the $1s$ state of the $\Xi^{-}$ hyperon in the $^{15}_{\Xi^{-}}$C hypernucleus, the IRRAWADDY and KINKA events have provided pertinent experimental data. However, constraining the $\Xi N$ interaction in theoretical models remains challenging due to inconsistencies in the experimental information. To minimize the impact of experimental uncertainties, this work incorporates the weighted average value of $B_{\Xi^{-}} = 6.46 \pm 0.25$ MeV from IRRAWADDY ($B_{\Xi^{-}} = 6.27 \pm 0.27$ MeV) and the larger KINKA values ($B_{\Xi^{-}} = 8.00 \pm 0.77$ MeV) as fitting targets. Furthermore, some research indicates the possibility of mixing between the $\Xi^{-}$ state in $^{14}$N and the $\Xi^{0}$ state in $^{14}$C within the $^{15}_{\Xi}$C event, potentially due to the $\Xi^{-}p \leftrightarrow \Xi^{0} n$ strong interaction charge exchange \cite{Friedman2021PLB820.136555, Friedman2023PLB837.137640}. For instance, the IRRAWADDY event has been interpreted as the $^{14}$C + $\Xi^{0}_{p}$ state in Ref. \cite{Friedman2023PLB837.137640}. Accordingly, the related event of $^{13}_{\Xi}$B can serve as another fitting target for constructing the $\Xi N$ interaction, as they can be interpreted as $^{12}$C + $\Xi^{-}_{p}$ without the need to account for $\Xi^{0}$ mixing \cite{Friedman2023PLB837.137640}.

\begin{table*}[hbpt]
  \centering
  \caption{The ratio of $\sigma$-$\Xi$ coupling strengths $g_{\sigma\Xi}/g_{\sigma N}$ for various RMF effective interactions, which are determined by fitting to the possible experimental values of the $\Xi^{-}$ separation energy of $^{15}_{\Xi^{-}}$C in the $1s$ state \cite{Yoshimoto2021PTEP2021.073D02} (denoted as $\Xi$C$s$), in the $1p$ state \cite{Hayakawa2021PRL126.062501} (denoted as $\Xi$C$p$), and of $^{13}_{\Xi^{-}}$B in the $1p$ state \cite{Aoki2009NPA828.191} (denoted as $\Xi$B$p$), see text for details. For other meson-hyperon coupling channels, the ratio of coupling strengthes are fixed to be $g_{\omega\Xi}/g_{\omega N}=0.333$, $g_{\rho\Xi}/g_{\rho N}=1.000$, $g_{\delta\Xi}/g_{\delta N}=1.000$, and additionally the $\omega$-$\Xi$ tensor coupling $f_{\omega\Xi}=-0.400g_{\omega\Xi}$.}\label{Tab:CouplingStrength}
  \renewcommand{\arraystretch}{1.25}
  \doublerulesep 0.1pt \tabcolsep 14pt
\begin{tabular}{cccccccc}
\hline\hline
                 & PK1      & TW99     & PKDD     & DD-ME2   & DD-MEX   & DD-ME$\delta$  & DD-LZ1   \\ \hline
$\Xi$C$s$        & 0.304666 & 0.309145 & 0.312701 & 0.313264 & 0.309712 & 0.319533       & 0.305429 \\
$\Xi$C$p$        & 0.312236 & 0.318984 & 0.321078 & 0.322175 & 0.320552 & 0.324708       & 0.322607 \\
$\Xi$B$p$        & 0.320842 & 0.326105 & 0.328357 & 0.329127 & 0.326959 & 0.332777       & 0.327859 \\
\hline\hline 
\end{tabular}
\end{table*}

For the selected RMF effective interactions, three different fitting strategies were employed to construct the $\Xi N$ interaction. Specifically, the first strategy involved fitting the separation energy of the hyperon $1s$ state in the $^{15}_{\Xi^{-}}$C hypernucleus to the weighted average of the IRRAWADDY and the larger KINKA values, $B_{\Xi^{-}} = 6.46$ MeV, resulting in a set of interactions labeled as $\Xi$C$s$. The second strategy fitted the separation energy of the hyperon $1p$ state in the $^{15}_{\Xi^{-}}$C hypernucleus to 1.13 MeV, yielding another set of interactions called $\Xi$C$p$. The third strategy fitted the separation energy of the hyperon $1p$ state in the $^{13}_{\Xi^{-}}$B hypernucleus to 0.82 MeV, producing a set referred to as $\Xi$B$p$, as detailed in Table \ref{Tab:CouplingStrength}. As the spin-orbit splitting feature of $\Xi^{-} {1p}$ state was not exactly distinguished in the experiment, the fitting for both the $\Xi$C$p$ and $\Xi$B$p$ series was conducted by averaging the values of the $\Xi^{-}$ spin doublet with the same orbital angular momentum $l_{\Xi^{-}}$. In previous studies, there have been discussions regarding the ``spurious" contributions arising from the $\rho$ meson in the self-energy of $\Xi^{-}$ hyperons, along with the corresponding details of their subtraction \cite{Jennings1994PRC49.2472}. In this work, the same methodology is applied to address this issue, and a similar treatment is employed for the isovector scalar $\delta$ meson. Additionally, considering the weak coupling between the $\Xi$ hyperon and nucleons, the bulk and single-particle properties of hypernuclei are sensitive to $\Xi N$ interactions. To achieve precise results, the coupling strength $g_{\sigma \Xi}/g_{\sigma N}$ is maintained to six decimal places. From Table \ref{Tab:CouplingStrength}, it is observed that the density-dependent RMF effective interaction yields a systematic increase in the coupling strength of the $ g_{\sigma\Xi}/g_{\sigma N} $ compared to PK1. Furthermore, for all the RMF models employed, the strength of the $\Xi$C$p$ effective interaction is generally slightly higher than that of the $\Xi$C$s$, and lower than that of the $\Xi$B$p$.

\subsection{$\Xi^{-}$ separation energies and hyperon local potential}
In this section, the $\Xi^{-}$ hyperon separation energies in the hypernuclei $^{15}_{\Xi^{-}}$C, $^{13}_{\Xi^{-}}$B, and $^{12}_{\Xi^{-}}$Be are calculated using the selected RMF models and the three sets of $\Xi N$ interactions listed in Table \ref{Tab:CouplingStrength}, with the $\Xi^{-}$ hyperon considered in either the $1s$ or $1p$ state (indicated by the index). Additionally, the $\Xi^{-}$ hyperon potential $U_{\Xi^{-}}(\rho_{0})$ in symmetric nuclear matter at saturation density is provided, as shown in Table \ref{Tab:SeparationEnergy} \cite{Tu2022APJ925.16}. The bolded values in the table correspond to the experimental data targeted during the fitting process for the $\Xi N$ interactions. From Table \ref{Tab:SeparationEnergy}, it is observed that the separation energies of hyperons increase gradually from $\Xi$C$s$ to $\Xi$C$p$ to $\Xi$B$p$ for the three $\Xi N$ interactions. For all selected RMF models, the difference between the results calculated using $\Xi$B$p$ and those using $\Xi$C$s$ is used to reflect the variation in hyperon separation energies predicted by the model, illustrating the so-called model dependence. Among these, the nonlinear effective interaction PK1 generally yields results that are more consistent with experimental or empirical data, as discussed in Ref. \cite{Sun2016PRC94.064319}. For density-dependent RMF effective interactions, the results exhibit significant model dependence. Specifically, the DD-LZ1 model, while reproducing one hyperon separation energy, often shows the greatest discrepancy from experimental values in other cases. After further consideration of the $\delta$ meson within the $\Xi N$ interaction, i.e. DD-ME$ \delta $, these discrepancies between models are significantly reduced, resulting in theoretical calculations that are more consistent with experimental observations. In the subsequent discussion, we will further analyze the reasons for the optimized results achieved by the additional introduction of the isovector scalar coupling channel.

\begin{table*}[hbpt]
	\centering
	\footnotesize
	\caption{The calculated $\Xi^{-}$ separation energies $B_{\Xi^{-}}$ (in MeV) of the hypernuclei $^{15}_{\Xi^{-}}$C, $^{13}_{\Xi^{-}}$B, and $^{12}_{\Xi^{-}}$Be by assuming $\Xi^{-}$ in the $1s$ or $1p$ state (marked by the index) with various RMF effective interactions listed in Table \ref{Tab:CouplingStrength}, along with the referred experimental data. Additionally, the $\Xi^{-}$ hyperon potential $U_{\Xi^{-}}(\rho_{0})$ in symmetric nuclear matter at saturation density was included.}\label{Tab:SeparationEnergy}
	\renewcommand{\arraystretch}{1.25}
	\doublerulesep 0.1pt \tabcolsep 10pt
	\begin{tabular}{ccccccccc}
		\hline\hline
		&                                    & $ ^{15}_{\Xi^{-}_{s}} $C                             & $ ^{15}_{\Xi^{-}_{p}} $C                          & $ ^{13}_{\Xi^{-}_{s}} $B & $ ^{13}_{\Xi^{-}_{p}} $B                   & $ ^{12}_{\Xi^{-}_{s}} $Be                                    & $U_{\Xi^{-}}(\rho_{0})$ \\ \hline
		\multirow{7}{*}{$\Xi$C$s$}   & PK1                                & $\bm{6.460}$                                         & 0.449                                             & 5.808                    & -0.505                                     & 3.463                                                        & -13.286                 \\
		& TW99                               & $\bm{6.460}$                                         & 0.155                                             & 5.524                    & -0.652                                     & 3.085                                                        & -14.311                 \\
		& PKDD                               & $\bm{6.460}$                                         & 0.304                                             & 5.590                    & -0.549                                     & 3.189                                                        & -13.946                 \\
		& DD-ME2                             & $\bm{6.460}$                                         & 0.290                                             & 5.793                    & -0.524                                     & 3.327                                                        & -14.384                 \\
		& DD-MEX                             & $\bm{6.460}$                                         & 0.111                                             & 5.955                    & -0.652                                     & 3.189                                                        & -14.533                 \\
		& DD-ME$ \delta $                    & $\bm{6.460}$                                         & 0.637                                             & 5.250                    & -0.288                                     & 3.382                                                        & -13.802                 \\
		& DD-LZ1                             & $\bm{6.460}$                                         & -0.120                                            & 7.206                    & -0.810                                     & 4.003                                                        & -13.010                 \\ \hline
		\multirow{7}{*}{$\Xi$C$p$}   & PK1                                & 8.495                                                & $\bm{1.130}$                                      & 8.025                    & -0.027                                     & 5.301                                                        & -16.092                 \\
		& TW99                               & 9.628                                                & $\bm{1.130}$                                      & 8.729                    & -0.002                                     & 5.903                                                        & -18.423                 \\
		& PKDD                               & 8.989                                                & $\bm{1.130}$                                      & 8.168                    & 0.022                                      & 5.377                                                        & -17.321                 \\
		& DD-ME2                             & 9.163                                                & $\bm{1.130}$                                      & 8.606                    & 0.071                                      & 5.809                                                        & -17.966                 \\
		& DD-MEX                             & 10.093                                               & $\bm{1.130}$                                      & 9.797                    & 0.063                                      & 6.637                                                        & -19.061                 \\
		& DD-ME$ \delta $                    & 7.716                                                & $\bm{1.130}$                                      & 6.437                    & 0.065                                      & 4.460                                                        & -15.703                 \\
		& DD-LZ1                             & 13.217                                               & $\bm{1.130}$                                      & 14.788                   & 0.168                                      & 11.168                                                       & -20.136                 \\ \hline
		\multirow{7}{*}{$\Xi$B$p$}   & PK1                                & 11.068                                               & 2.173                                             & 10.887                   & $\bm{0.820}$                               & 7.766                                                        & -19.282                 \\
		& TW99                               & 12.128                                               & 2.161                                             & 11.295                   & $\bm{0.820}$                               & 8.263                                                        & -21.399                 \\
		& PKDD                               & 11.386                                               & 2.124                                             & 10.649                   & $\bm{0.820}$                               & 7.578                                                        & -20.254                 \\
		& DD-ME2                             & 11.461                                               & 2.048                                             & 11.026                   & $\bm{0.820}$                               & 8.032                                                        & -20.761                 \\
		& DD-MEX                             & 12.434                                               & 2.041                                             & 12.296                   & $\bm{0.820}$                               & 8.977                                                        & -21.737                 \\
		& DD-ME$ \delta $                    & 9.817                                                & 2.088                                             & 8.449                    & $\bm{0.820}$                               & 6.332                                                        & -18.668                 \\
		& DD-LZ1                             & 15.497                                               & 1.818                                             & 17.318                   & $\bm{0.820}$                               & 13.601                                                       & -22.315                 \\ \hline
		\multicolumn{2}{c}{\multirow{3}{*}{\rm{Expt. or empirical data}}} & $ 6.46 \pm 0.25$ \cite{Yoshimoto2021PTEP2021.073D02} & $ 1.13 \pm 0.14$ \cite{Hayakawa2021PRL126.062501} &                          & $ 0.82 \pm 0.17$ \cite{Aoki2009NPA828.191} & 4.50 \cite{Khaustov2000PRC61.054603, Hiyama2008PRC78.054316} &                         \\
		&                                    &                                                      &                                                   &                          &                                            & 6.30 \cite{Nagae2019AIP.Conf.Proc.2130.020015}               &                         \\
		&                                    &                                                      &                                                   &                          &                                            & 9.00(2.00) \cite{Nagae2019AIP.Conf.Proc.2130.020015}         &                         \\ \hline\hline
		
	\end{tabular}
\end{table*}

Further examination of the $\Xi$C$s$ and $\Xi$C$p$ results in Table \ref{Tab:SeparationEnergy} reveals that the predicted $ B_{\Xi^{-}} $ for $^{13}_{\Xi^{-}_{p}}$B may be negative across various effective Lagrangians. As shown in Table \ref{Tab:CouplingStrength}, the meson-hyperon coupling strengths $g_{\sigma\Xi}/g_{\sigma N}$ for $\Xi$C$s$ and $\Xi$C$p$ are generally lower than those for $\Xi$B$p$, which weakens the attractive contributions from the meson fields and correspondingly the Coulomb field between $\Xi^{-}$ and protons, leading to more weakly bound results. In fact, due to their small $\Xi N$ coupling strengths, all RMF-$\Xi$C$s$ models predict negative hyperon separation energies for $^{13}_{\Xi^{-}_{p}}$B, which shows a discrepancy from the experimental value $B_{\Xi^{-}}=0.82$ MeV. One possible explanation for this deviation is attributed to the effect of deformation. Previous studies based on the axially deformed Skyrme-Hartree-Fock model suggest that incorporating deformation effects might lead to a more consistent description between theory and experiment \cite{Guo2021PRC104.L061307}. The consideration of deformation is expected to bring corrections to the separation energy ranging from $0.54$ to $0.98$ MeV. Since this work focuses primarily on the treatment of nuclear in-medium effects and the impact of isovector scalar $ \delta $ meson on the bulk and single-particle properties of hypernuclei, deformation effects are not considered in the current models. Notably, except for $^{13}_{\Xi^{-}_{p}}$B, the results from DD-ME$\delta$-$\Xi$C$s$ are generally consistent with those from the density-dependent optical potential methodology \cite{Friedman2022EPJWebConf.271.03002, Friedman2023PLB837.137640}.

\begin{table*}[hbpt]
	\centering
	\caption{The binding energies $E$ (in MeV) of the hypernuclei $^{15}_{\Xi^{-}}$C and $^{15}_{\Xi^{0}}$C, assuming $\Xi$ is in the $1s$ or $1p$ state (marked by the index), as well as their nucleonic cores $^{14}$N and $^{14}$C, are calculated using the DD-ME$\delta$-$\Xi$C$p$ and DD-ME$\delta$-$\Xi$B$p$ interactions. Additionally, the hyperon separation energy $B_{\Xi^{-}_{s}}$ is provided, along with the energy difference $B_{\Xi^{0}_{p}}^{*}$ between the near-threshold $^{14}$C+$\Xi^{0}_{p}$ ($^{15}_{\Xi^{0}_{p}}$C) nuclear state and the $^{14}$N+$\Xi^{-}$ threshold.}\label{Tab:Bindingenergy}
	\renewcommand{\arraystretch}{1.25}
	\doublerulesep 0.1pt \tabcolsep 12 pt
	\begin{tabular}{ccccccccc}
		\hline\hline
		DD-ME$\delta$ & \multicolumn{1}{c}{$^{14}$N} & \multicolumn{1}{c}{$^{15}_{\Xi^{-}_{s}}$C} & \multicolumn{1}{c}{$^{15}_{\Xi^{-}_{p}}$C} & $B_{\Xi^{-}_{s}}$ & $^{14}$C                  & \multicolumn{1}{c}{$^{15}_{\Xi^{0}_{s}}$C} & \multicolumn{1}{c}{$^{15}_{\Xi^{0}_{p}}$C}  & $B_{\Xi^{0}_{p}}^{*}$  \\ \hline
		$\Xi$C$p$     & \multirow{2}{*}{-104.679}    & -112.395                                   & -105.809                                   & 7.716             & \multirow{2}{*}{-105.720} & -112.025                                   & -105.284                                    & 6.161                  \\
		$\Xi$B$p$     &                              & -114.496                                   & -106.768                                   & 9.817             &                           & -114.108                                   & -106.320                                    & 7.197                  \\ \hline\hline
	\end{tabular}
\end{table*}

\begin{figure}[htbp]
	\centering
	\includegraphics[width=0.48\textwidth]{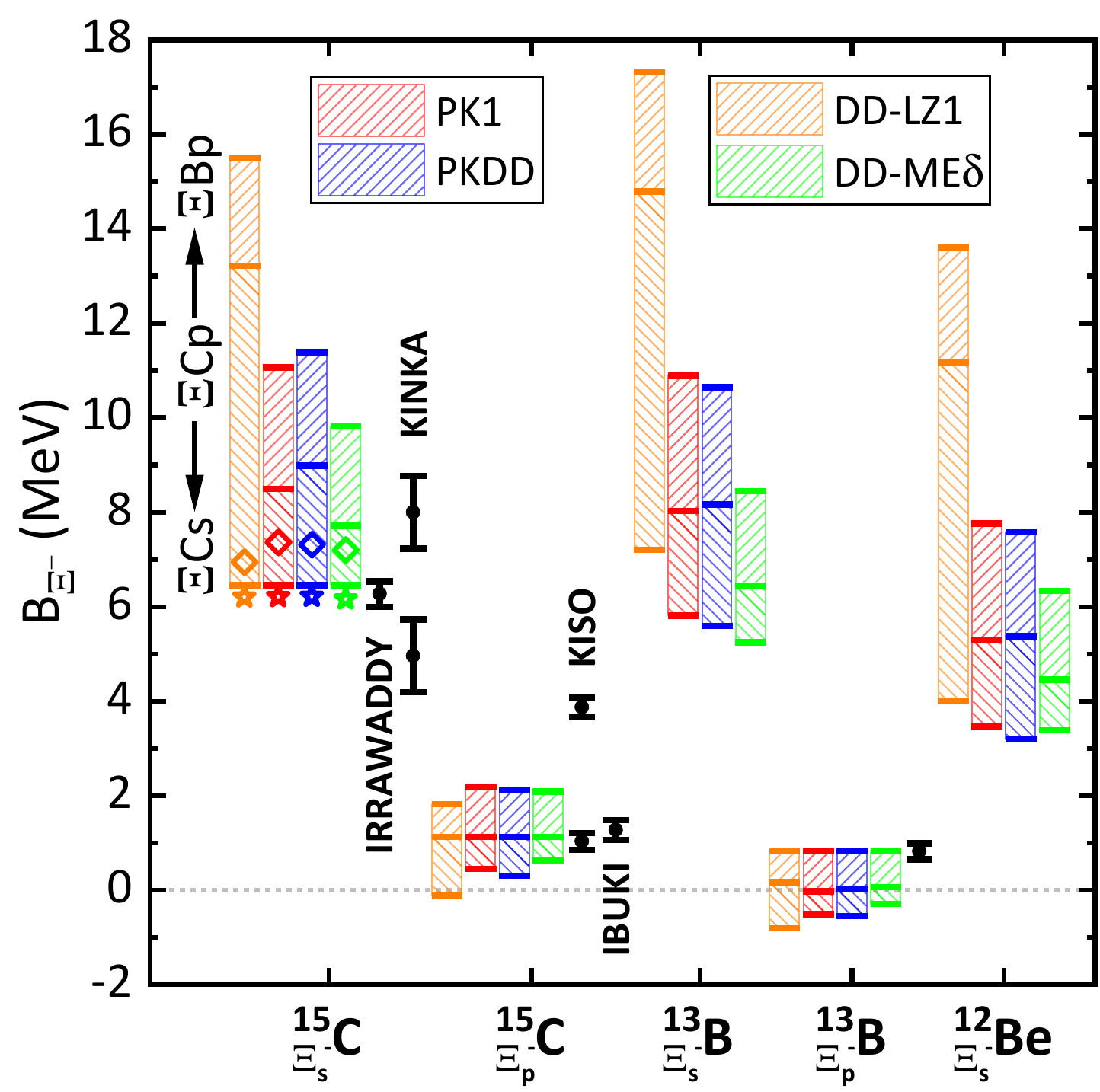}
	\caption{The calculated $\Xi^{-}$ separation energies $B_{\Xi^{-}}$ of the hypernuclei $^{15}_{\Xi^{-}}$C, $^{13}_{\Xi^{-}}$B, and $^{12}_{\Xi^{-}}$Be were obtained by assuming $\Xi^{-}$ in the $1s$ or $1p$ state with RMF effective interactions PK1, PKDD, DD-LZ1 and DD-ME-$\delta$. The experimental data for $\Xi^{-}$ hypernucleus are marked with black error bars. The left-slash patterns represent the differences in results based on $\Xi$B$p$ and $\Xi$C$p$, while the right-slash patterns illustrate the differences between results from $\Xi$C$p$ and $\Xi$C$s$. Additionally, the energy difference $ B_{\Xi^{0}_{p}}^{*} $ between the $^{14}$C+$\Xi^{0}_{p}$ nuclear bound state and the $^{14}$N+$\Xi^{-}$ threshold is also shown, given by the effective interactions $\Xi$B$p$ (marked by diamonds) and $\Xi$C$p$ (marked by stars), respectively.}\label{Fig:SeparationEnergySum}
\end{figure}

As seen in Table \ref{Tab:SeparationEnergy}, another sizable difference appears in the hyperon separation energy of $^{15}_{\Xi^{-}_{s}}$C given by the $\Xi$C$p$ and $\Xi$B$p$ models and the experimental data. A possible explanation for this discrepancy may involve theoretical considerations of $\Xi^{-}$ conversion to $\Xi^{0}$ or the mixing of $\Xi^{0}$ within $^{15}_{\Xi}$C, in an effort to align theoretical predictions with experimental observations \cite{Friedman2023PLB837.137640}. In that study, the identification of the $\Xi^{-}_{s}$ nuclear bound state with IRRAWADY is questioned, and an alternative assignment as a near-threshold $^{14}$C+$\Xi^{0}_{p}$ nuclear bound state is suggested, with a reported threshold energy of $6.17 \pm 0.21$ MeV relative to $^{14}$N+$\Xi^{-}$. Following this idea, taking the effective Lagrangian DD-ME$\delta$ as an example, the binding energies $E$ of the hypernuclei $^{15}_{\Xi^{-}}$C and $^{15}_{\Xi^{0}}$C, as well as their nucleonic cores $^{14}$N and $^{14}$C, are calculated based on the $\Xi$C$p$ and $\Xi$B$p$ interactions. For comparison, the energy difference $B_{\Xi^{0}_{p}}^{*}$ between the near-threshold $^{14}$C+$\Xi^{0}_{p}$ ($^{15}_{\Xi^{0}_{p}}$C) nuclear state and the threshold for $^{14}$N+$\Xi^{-}$ is also provided, as shown in Table \ref{Tab:Bindingenergy} as well as in Fig. \ref{Fig:SeparationEnergySum}. In the current calculations, $B_{\Xi^{0}_{p}}^{*}$ can be expressed as follows
\begin{align}\label{eq:threshold}
	B_{\Xi^{0}_{p}}^{*}= (m_{p} - m_{n}) + \left[E(^{14}\mathrm{N}) - E(^{15}_{\Xi^{0}_{p}}\mathrm{C})\right] + (m_{\Xi^{-}} - m_{\Xi^{0}}),
\end{align}
where the masses of bare nucleons and hyperons are taken from Ref. \cite{Navas2024PRD110.030001}. To facilitate the comparison between the states of $\Xi^{-}$ and $\Xi^{0}$, the hyperon separation energy $B_{\Xi^{-}_{s}}$ of $^{15}_{\Xi^{-}_{s}}$C is also given. It is seen that the values of $B_{\Xi^{0}_{p}}^{*}$ are 6.161 MeV for $\Xi$C$p$ and 7.197 MeV for $\Xi$B$p$, respectively, which are much closer to the weighted average of $6.46 \pm 0.25$ MeV from IRRAWADDY and the larger KINKA values as compared to $B_{\Xi^{-}_{s}}$, and are also in agreement with the referred result in Ref. \cite{Friedman2023PLB837.137640}.

To clarify the impact of model selection on the results, four typical RMF Lagrangians PK1, PKDD, DD-LZ1, and DD-ME$\delta$ were selected to illustrate the differences in hyperon separation energies under various $\Xi N$ interactions and their comparison with experimental data, as shown in Fig. \ref{Fig:SeparationEnergySum}. The left-slash pattern represents the differences based on $\Xi$B$p$ and $\Xi$C$p$ results, while the right-slash pattern denotes the differences between $\Xi$C$p$ and $\Xi$C$s$. Black error bars indicate the existing experimental data of the selected hypernuclei. As shown in Table \ref{Tab:SeparationEnergy}, among the selected RMF Lagrangians, DD-LZ1 displays the most significant model dependence. The density-dependent effective Lagrangian PKDD shows a significant reduction in the differences among various $\Xi N$ interactions, aligning closely with the nonlinear effective Lagrangian PK1. By considering additional meson-baryon degrees of freedom, namely by including the isovector scalar $\delta$ meson, the model discrepancy with DD-ME$\delta$ is further reduced. Although it still cannot fully reproduce all experimental data, the more reasonable treatment of nuclear in-medium effects and the more comprehensive consideration of meson degrees of freedom clearly have significant implications for reducing model dependence and for looking into the internal structure of hypernuclei, and are worthy of our in-depth exploration.

\begin{figure}[htbp]
	\centering
	\includegraphics[width=0.48\textwidth]{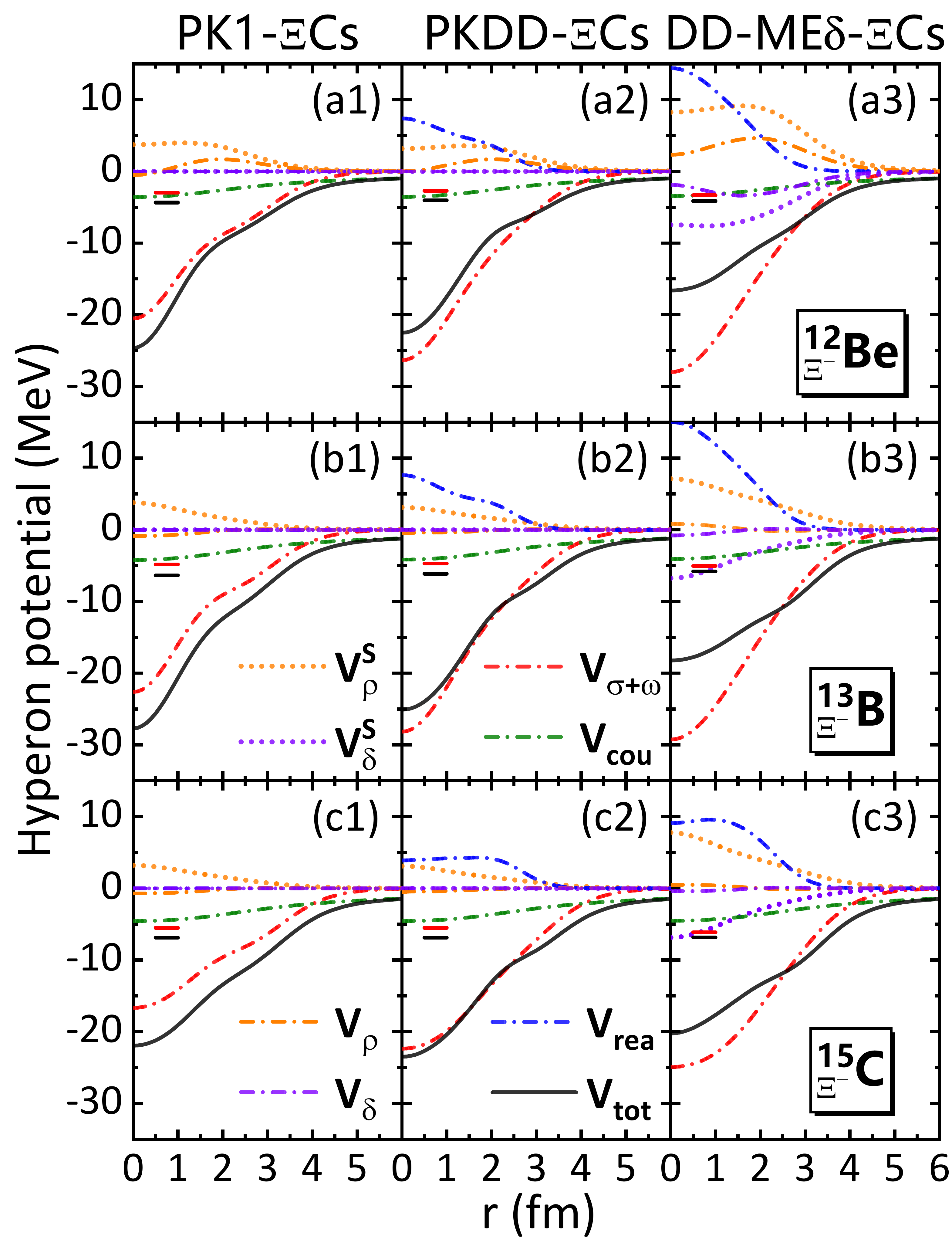}
	\caption{Local $\Xi^{-}$ mean-field potentials (solid curves) in $ ^{12}_{\Xi^{-}} $Be, $ ^{13}_{\Xi^{-}} $B and $ ^{15}_{\Xi^{-}} $C, decomposed by their contributions (dash-dotted lines) from various mesons ($V_{\sigma+\omega}$,~$V_{\rho}$,~$V_{\delta}$) and photon ($V_{\rm{cou}}$) channels as well as the rearrangement terms ($V_{\rm{rea}}$), calculated by the RMF effective interactions PK1-$\Xi $C$ s $, PKDD-$\Xi $C$ s $ and DD-ME$\delta$-$\Xi $C$ s $. For $\rho$ and $\delta$ mesons, the dotted lines denote the ``spurious" results without removing the contribution due to the hyperon interacting with itself, see text for details. In addition, the $\Xi^{-}1s_{1/2}$ single-particle energies are shown by the black levels compared with their ``spurious" results by the red levels.}\label{Fig:Potential}
\end{figure}

To investigate the impact on the description of hyperon separation energy in light hypernuclei, we selected the nonlinear effective interaction PK1-$\Xi$C$s$, the density-dependent effective interaction PKDD-$\Xi$C$s$ and DD-ME$ \delta $-$\Xi$C$s$. By utilizing these interactions, we performed calculations on the hyperon's local self-energy $\Sigma^{\Xi^{-}}_{+}$ in $^{12}_{\Xi^{-}}$Be, $^{13}_{\Xi^{-}}$B, and $^{15}_{\Xi^{-}}$C, incorporating contributions from various mesons and the photon, as shown in Fig. \ref{Fig:Potential}. For simplicity in notation, we refer to $\Sigma^{+}_{\Xi^{-}}$ as $V_{\rm{tot}}$ in the figures and subsequent discussions. The figures show results with and without removing the ``spurious" contributions from the hyperon self-energy, indicated by $V_{\rho}^{\rm{S}}$ ($V_{\delta}^{\rm{S}}$) and $V_{\rho}$ ($V_{\delta}$), respectively. Accordingly, ($V_{\rho}^{\rm{S}}$-$V_{\rho}$) and ($V_{\delta}^{\rm{S}}$-$V_{\delta}$) represent the ``spurious" contributions arising from the self-interaction of hyperons due to isovector mesons $ \rho $ and $ \delta $, respectively, which are removed in the calculations.

Subsequently, the differences in separation energy results arising from three sets of effective interactions are further elucidated from the perspective of potential. Specifically, for $^{15}_{\Xi^{-}}$C, as shown in Fig. \ref{Fig:Potential}(c1)-(c3), the contributions from isoscalar mesons $\sigma$ and $\omega$ are dominant among the selected effective interactions. Notably, the density-dependent effective interactions generally yield deeper potentials $V_{\sigma+\omega}$ (red lines) compared to PK1-$\Xi$C$s$. Given that $^{15}_{\Xi^{-}}$C has an $N=Z$ core, the contributions from isovector mesons $V_{\rho}$ and $V_{\delta}$ are approximately negligible. Therefore, for the nonlinear effective interaction PK1, the hyperon potential is approximated by the sum of the contributions from the isoscalar meson $V_{\sigma+\omega}$ and the photon $V_{\rm{cou}}$. By comparison, the density-dependent effective interactions require additional consideration of the contribution of the rearrangement term $V_{\rm{rea}}$ due to the density dependence of the meson-baryon coupling strengths. This significant repulsive contribution $V_{\rm{rea}}$ (blue lines) counters the Coulomb attractive contribution $V_{\rm{cou}}$ (olive lines) of the photon, resulting in a hyperon potential (black lines) from the density-dependent effective interaction that is similar to or even shallower than those from PK1-$\Xi$C$s$ at the center. For DD-ME$\delta$-$\Xi$C$s$, compared to PKDD-$\Xi$C$s$, there is a significant increase in the central contribution from the rearrangement terms, which can be attributed to the additional inclusion of the isovector scalar $\delta$ meson. Additionally, the repulsive contribution from the rearrangement terms decays rapidly with increasing radial radius. As a result, a wider and deeper hyperon potential is obtained at the nuclear surface. For the hyperon occupying the $1p$ orbital, which is mainly distributed near the surface of the hypernucleus, this potential provides sufficient binding. Consequently, the model offers a theoretical description of the separation energies for both the $1s$ and $1p$ hyperon states that aligns more closely with experimental observations.

Similar phenomena are observed for other light hypernuclei such as $ ^{12}_{\Xi^{-}} $Be and $ ^{13}_{\Xi^{-}} $B, as shown in Fig. \ref{Fig:Potential}(a1)-(a3) and Fig. \ref{Fig:Potential}(b1)-(b3). Although DD-ME$ \delta $-$\Xi$C$s$ yields the deepest $ V_{\sigma+\omega} $, it is largely counteracted by the strong repulsion at the center, resulting in the shallowest potential. Conversely, the nonlinear effective interaction PK1-$\Xi$C$s$, where the rearrangement term contribution is zero, often produces the deepest potential, leading to the largest $ 1s $ state separation energy for $ ^{12}_{\Xi^{-}} $Be and $ ^{13}_{\Xi^{-}} $B. As an extension, we also compared the impact of the ``spurious" contributions from isovector mesons in hyperon self-interactions on the hyperon single-particle energies, as illustrated by the dashed lines in Fig. \ref{Fig:Potential}. The black and red dashed lines represent the $1s_{1/2}$ state energy of $\Xi^{-}$ hyperons both with and without the removal of ``spurious" contributions. When considering the contributions of isovector mesons in the hyperon potential, we observe that without removing the ``spurious" contributions in hyperon self-interactions, both effective interactions PK1-$\Xi$C$s$ and PKDD-$\Xi$C$s$ predict larger $V_{\rho}^{\rm{S}}$, indicating a significant influence of the isovector meson on the single-particle energies. In contrast, although DD-ME$\delta$-$\Xi$C$s$ exhibits a larger $V_{\rho}^{\rm{S}}$, its effect is largely offset by $V_{\delta}^{\rm{S}}$, making its single-particle energies being less sensitive to the treatment of isovector meson.

\subsection{Systematics of  $\Xi^{-}$  hypernuclear properties}
\begin{figure}[htbp]
  \centering
  \includegraphics[width=0.48\textwidth]{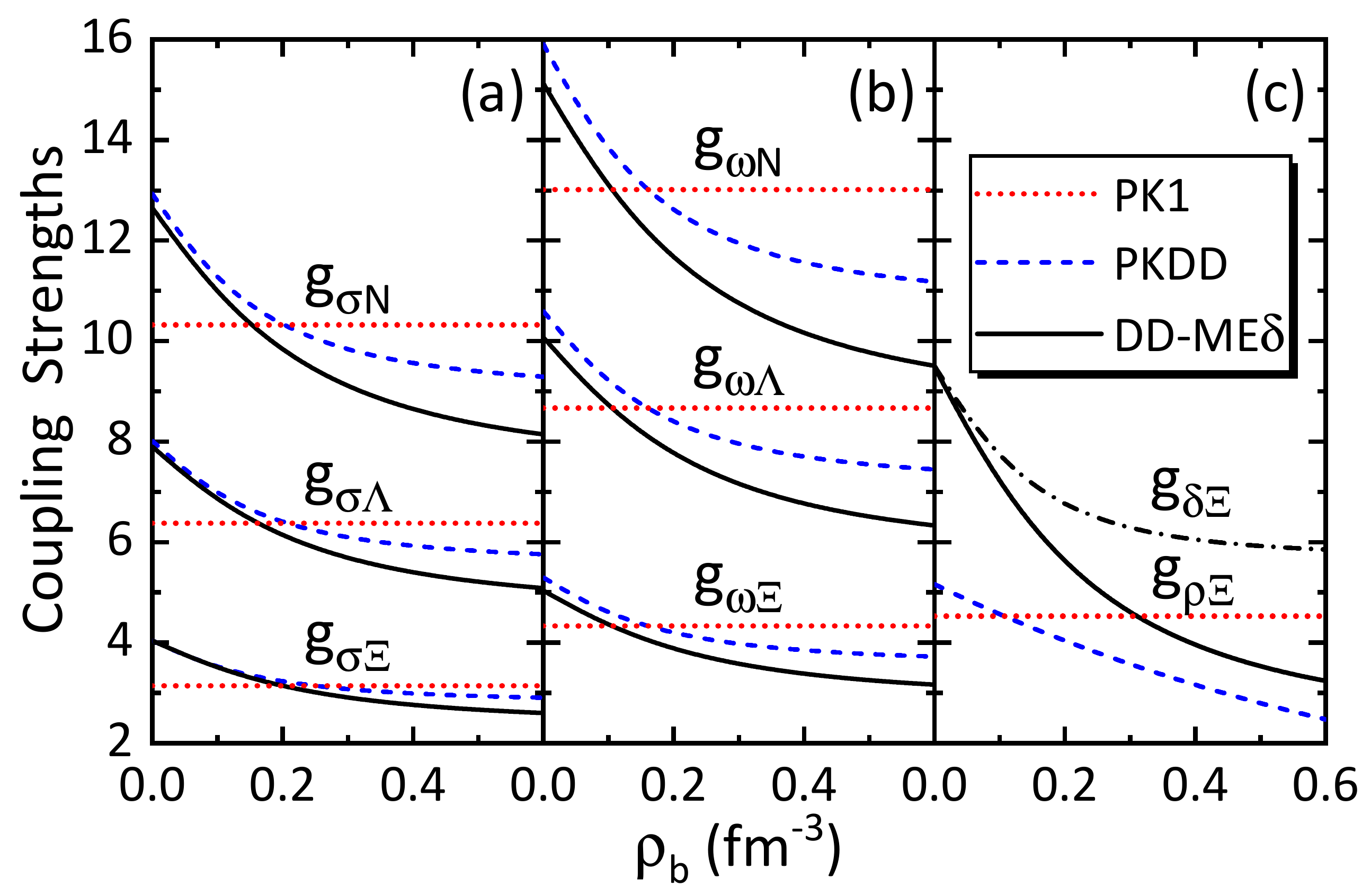}
  \caption{Meson-nucleon/$\Xi$ coupling strengths as a function of baryonic density $ \rho_{b} $ within the RMF effective interactions PK1, PKDD and DD-ME$ \delta $, including the isoscalar $ g_{\sigma B} $ and $ g_{\omega B} $ [panel (a) and (b)] and the isovector $ g_{\rho B} $ and $ g_{\delta B} $ [panel (c)]. As a comparison, the meson-$\Lambda$ couplings are given as well, with their values taken from Ref. \cite{Ding2023CPC47.124103} for PK1 and PKDD, and Ref. \cite{Tu2022APJ925.16} for DD-ME$ \delta $.}\label{Fig:CouplingStrength}
\end{figure}

To understand the significant differences in the evolution of contributions from rearrangement terms in the hyperon potential as a function of density, the meson-nucleon/hyperon coupling strengths for three selected sets of effective interactions are presented in Fig. \ref{Fig:CouplingStrength}(a), (b), and (c), corresponding respectively to the isoscalar scalar channel $ g_{\sigma B} $, isoscalar vector channel $ g_{\omega B} $, and isovector channels $ g_{\rho B} $ and $ g_{\delta B} $. For reference, the results for the $ \Lambda $ hyperon are also provided. Compared to PKDD, in DD-ME$ \delta $, the meson-baryon coupling strength exhibits stronger density dependence, particularly evident in the isovector channel. These differences arise primarily from the distinct forms of density dependence for the isovector mesons, as detailed in Eqs. \eqref{eq:coupling_constants1} and \eqref{eq:coupling_constants2}. Additionally, DD-ME$\delta$ incorporates the $\delta$ meson, whose coupling strength is sensitive to density as shown by the black dashed line, and contributes significantly more to the rearrangement compared to PKDD. Notably, while the coupling strength in the isovector channel of DD-ME$\delta$ is roughly twice that of PKDD at low densities, it diminishes rapidly with increasing baryon density. This rapid decline in coupling strength in the isovector channel is the primary reason for the pronounced reduction in rearrangement contributions from the center to the surface in DD-ME$\delta$, as illustrated in Fig. \ref{Fig:Potential}.

\begin{figure}[htbp]
  \centering
  \includegraphics[width=0.48\textwidth]{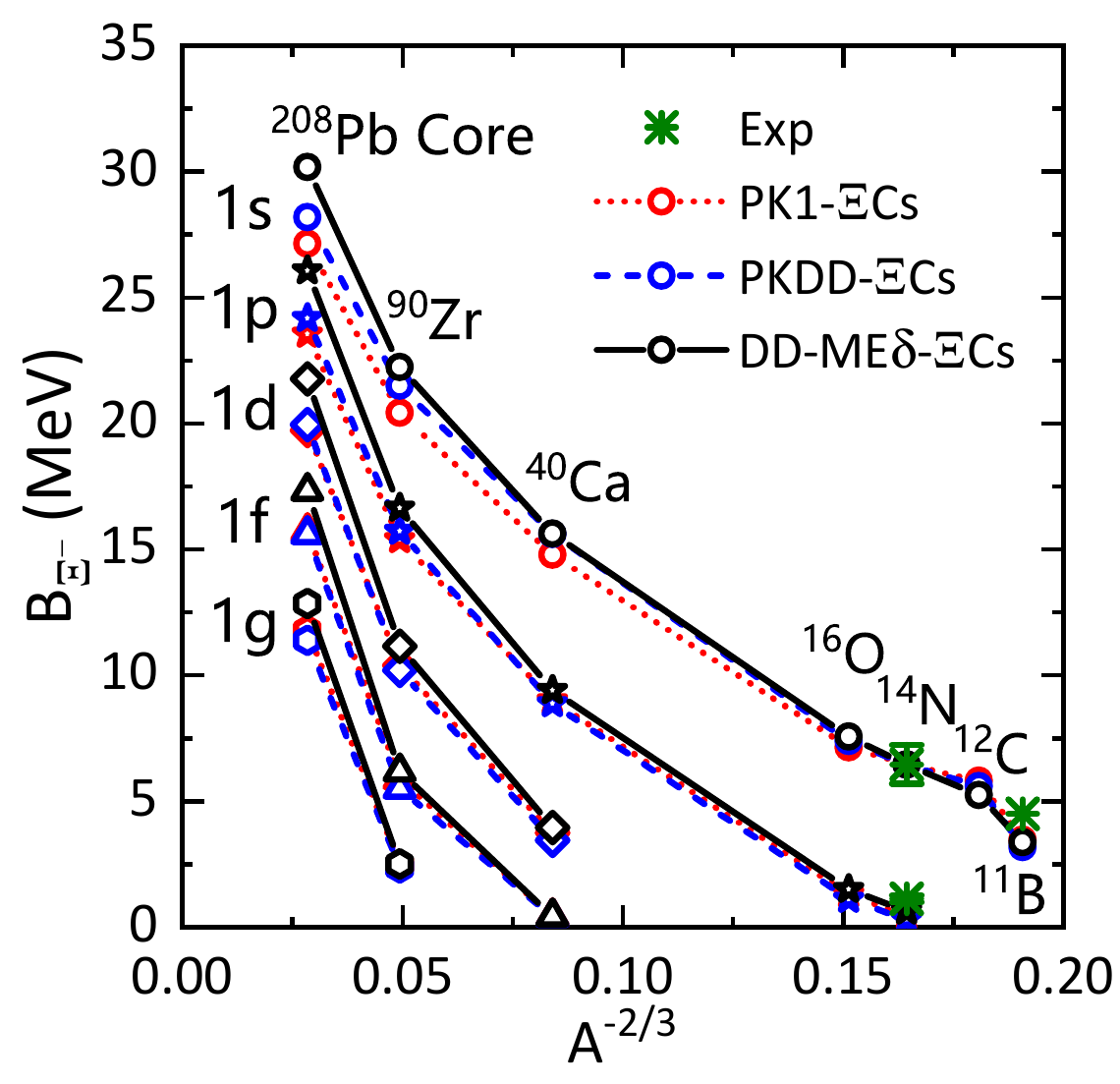}
	\caption{The calculated $\Xi^{-}$ separation energies $ B_{\Xi^{-}} $ for the single-$\Xi^{-}$ hypernuclei (labeled by their nucleonic cores) with the $\Xi$C$s$ meson-hyperon effective interactions in three types of RMF Lagrangians. For comparison, the experimental data taken from Refs. \cite{Khaustov2000PRC61.054603, Hiyama2008PRC78.054316, Hayakawa2021PRL126.062501, Yoshimoto2021PTEP2021.073D02} are also given.}\label{Fig:SeparationEnergy}
\end{figure}

To provide a more comprehensive understanding of the bulk and single-particle properties of $\Xi^{-}$ hypernuclei, based on the $\Xi$C$s$ interaction presented in Table \ref{Tab:SeparationEnergy}, the hyperon separation energies in the single-$\Xi^{-}$ hypernuclei from $^{12}_{\Xi^{-}}$Be to $^{209}_{\Xi^{-}}$Tl are systematically calculated and illustrated in Fig. \ref{Fig:SeparationEnergy}. Note that the $ B_{\Xi^{-}} $ for the $ 1p $, $ 1d $, $ 1f $, and $ 1g $ orbits are determined by taking the average of the spin doublets. Considerable model dependency appears in the descriptions of the separation energies when going for large mass region, even though the $ \Xi N $ interactions are fitted with the same light hypernucleus $ ^{15}_{\Xi^{-}} $C. Among the three sets of effective interactions, DD-ME$ \delta $ exhibits the most pronounced variation with mass number.

\begin{table*}[hbpt]
	\centering
	\caption{The single-particle and bulk properties in different single-$\Xi^{-}$ hypernuclei and their nucleonic cores using density-dependent RMF effective interaction DD-ME$ \delta $-$ \Xi $C$ s $, including the single-particle energies $\varepsilon_{\rm{s.p.}}$, binding energies $E$, charge radii $R_{\rm{c}}$, hyperon radii $R_{\Xi^{-}}$, and hypernuclear matter radii $R_{\rm{m}}$.}\label{Tab:HypernucleiProperties}
	\setlength{\tabcolsep}{18pt}
	\renewcommand{\arraystretch}{1.25}
  \begin{tabular}{ccccccc}
  \hline\hline
	Nucleus &
	$\Xi^{-}(nlj)$ &
	$\varepsilon_{\rm{s.p.}}$(MeV) &
	$E$(MeV) &
	$R_{\rm{c}}$(fm) &
	$R_{\Xi^{-}}$(fm) &
	$R_{\rm{m}}$(fm) \\ \hline
	\multirow{1}{*}{$^{11}$B}             &       -              &     -    &  -72.199  &  2.474   &   -    & 2.381    \\
	\multirow{3}{*}{$^{12}_{\Xi^{-}}$Be}  & \textbf{$1s_{1/2}$}  &  -4.140  &  -75.580  &  2.456   & 3.076  & 2.442    \\
										  & \textbf{$1p_{1/2}$}  &  -0.054  &  -71.143  &  2.476   & 9.388  & 3.548    \\
										  & \textbf{$1p_{3/2}$}  &  -0.099  &  -71.209  &  2.476   & 9.029  & 3.471    \\ \hline
	\multirow{1}{*}{$^{12}$C}             &       -              &     -    &  -85.624  &  2.525   &   -    & 2.383    \\
	\multirow{3}{*}{$^{13}_{\Xi^{-}}$B}   & \textbf{$1s_{1/2}$}  &  -5.813  &  -90.874  &  2.504   & 2.820  & 2.412    \\
										  & \textbf{$1p_{1/2}$}  &  -0.502  &  -85.260  &  2.528   & 7.075  & 3.017    \\
										  & \textbf{$1p_{3/2}$}  &  -0.601  &  -85.412  &  2.528   & 6.578  & 2.929    \\ \hline
	\multirow{1}{*}{$^{14}$N}             &       -              &     -    & -104.679  &  2.627   &   -    & 2.490    \\
	\multirow{3}{*}{$^{15}_{\Xi^{-}}$C}   & \textbf{$1s_{1/2}$}  &  -6.841  & -111.139  &  2.605   & 2.774  & 2.500    \\
										  & \textbf{$1p_{1/2}$}  &  -1.087  & -105.220  &  2.629   & 5.607  & 2.809    \\
										  & \textbf{$1p_{3/2}$}  &  -1.236  & -105.413  &  2.629   & 5.284  & 2.766    \\ \hline
	\multirow{1}{*}{$^{16}$O}             &       -              &     -    & -130.201  &  2.689   &   -    & 2.555    \\
	\multirow{3}{*}{$^{17}_{\Xi^{-}}$N}   & \textbf{$1s_{1/2}$}  &  -7.848  & -137.772  &  2.667   & 2.732  & 2.555    \\
										  & \textbf{$1p_{1/2}$}  &  -1.756  & -131.592  &  2.691   & 4.843  & 2.744    \\
										  & \textbf{$1p_{3/2}$}  &  -1.938  & -131.808  &  2.690   & 4.648  & 2.723    \\ \hline
	\multirow{1}{*}{$^{40}$Ca}            &       -              &     -    & -345.728  &  3.423   &   -    & 3.304    \\
	\multirow{3}{*}{$^{41}_{\Xi^{-}}$K}   & \textbf{$1s_{1/2}$}  & -15.639  & -361.359  &  3.401   & 2.749  & 3.280    \\
										  & \textbf{$1p_{1/2}$}  &  -9.200  & -355.025  &  3.415   & 3.744  & 3.312    \\
										  & \textbf{$1p_{3/2}$}  &  -9.398  & -355.225  &  3.415   & 3.739  & 3.311    \\ \hline
	\multirow{1}{*}{$^{208}$Pb}           &       -              &     -    &-1633.296  &  5.503   &   -    & 5.565    \\
	\multirow{3}{*}{$^{209}_{\Xi^{-}}$Tl} & \textbf{$1s_{1/2}$}  & -30.216  &-1663.470  &  5.489   & 3.417  & 5.547    \\
										  & \textbf{$1p_{1/2}$}  & -26.012  &-1659.356  &  5.493   & 4.315  & 5.553    \\
										  & \textbf{$1p_{3/2}$}  & -26.033  &-1659.377  &  5.493   & 4.329  & 5.553    \\
   \hline\hline
  \end{tabular}
\end{table*}

Furthermore, in Table \ref{Tab:HypernucleiProperties} we employ the effective interaction DD-ME$\delta$-$\Xi$C$s$ to study the bulk and single-particle properties such as the single-particle energies, binding energies, and corresponding characteristic radii for the $\Xi^{-}$ hypernuclei and their nucleonic cores, with the $ \Xi^{-} $ hyperon occupying the $ 1s $ or $ 1p $ orbitals. It is noteworthy that when the hyperon occupies the $ 1s $ state, in lighter hypernuclei, the hyperon radius is slightly larger than the nuclear matter radius of the hypernucleus, which is consistent with the conclusions in Ref. \cite{Isaka2024PRC109.044317}. As the mass number increases, the hyperon radius initially decreases and then increases, which may be due to the competing effects of the Coulomb interaction and strong interaction between hyperons and nucleons. Additionally, when the hyperon occupies the $ 1p $ state, the spatial distribution of the hyperon becomes more diffuse, resulting in a further increase in the hyperon radius compared to when it is in the $ 1s $ state.

\section{Summary}\label{Summary and Outlook}

To consider the nuclear in-medium effects and underscore the significance of contributions from various meson-baryon coupling channels in hyperon-nucleon (hyperon-hyperon) interactions for describing the $ \Xi^{-} $ hypernuclear structure, the DDRMF theory was extended to include $ \Xi^{-} $ hyperon degrees of freedom. By fitting the experimental separation energies for the $ 1s $ and $ 1p $ states in $^{15}_{\Xi^{-}}$C and the $ 1p $ state in $^{13}_{\Xi^{-}}$B, three sets of $ \Xi N $ effective interactions, $\Xi$C$s$, $\Xi$C$p$ and $\Xi$B$p$, were derived. Based on these three sets of $ \Xi N $ effective interactions, the hyperon $ 1s $ state separation energies for $^{12}_{\Xi^{-}}$Be, $^{13}_{\Xi^{-}}$B, and $^{15}_{\Xi^{-}}$C, as well as the $ 1p $ state hyperon separation energies for $^{15}_{\Xi^{-}}$C and $^{13}_{\Xi^{-}}$B, were calculated. Since a possible mixing mechanism could exist between $\Xi^{-}$ states in $^{14}$N and $\Xi^{0}$ states in $^{14}$C when extracting data from the IRRAWADY experiment, the energy difference between $^{14}$C+$\Xi^{0}_{p}$ nuclear bound state and $^{14}$N+$\Xi^{-}$ threshold is also calculated, which is shown to be close to the experimental data and align with those reported in Ref. \cite{Friedman2023PLB837.137640}. It is then checked that the results from RMF effective interactions behave significant model dependence. When further considering more comprehensive meson-baryon coupling contributions, specifically by introducing the isovector scalar $\delta$ meson, the differences between the results obtained using DD-ME$\delta$ under different $\Xi N$ effective interactions are further reduced, leading to results that were more consistent with experimental observations.

Three sets of effective interactions, namely PK1-$\Xi$C$s$, PKDD-$\Xi$C$s$, and DD-ME$\delta$-$\Xi$C$s$, are selected to further investigate the influence of in-medium effects and the isovector scalar $ \delta $ meson on hyperon properties. By comparing the hyperon potentials and their decomposed contributions in light hypernuclei, it is found that the density-dependent effective interactions introduce an additional rearrangement term, enhancing the hyperon potential in the central region at lower densities. The inclusion of the isovector scalar $ \delta $ meson and the different treatments of the density-dependent isovector coupling strength cause DD-ME$ \delta $-$\Xi$C$s$ to exhibit more pronounced rearrangement term contributions compared to PKDD-$\Xi$C$s$. Additionally, in DD-ME$ \delta $, the rearrangement term's contribution rapidly diminishes with increasing baryon density, resulting in a relatively broader hyperon potential. Thus, DD-ME$ \delta $ provides a reasonable description of the $ 1s $ and $ 1p $ states of $^{15}_{\Xi^{-}}$C and yields smaller separation energies for $^{12}_{\Xi^{-}_{s}}$Be and $^{13}_{\Xi^{-}_{s}}$B due to its relatively shallow hyperon potential at the center.

Subsequently, based on the $\Xi$C$s$ effective interaction, the separation energies of $\Xi^{-}$ hypernuclei from light to heavy are systematically calculated. The results reveal that the model dependence of the hyperon separation energies obtained from the $ \Xi N $ interaction, fitted with the $^{15}_{\Xi^{-}}$C hypernuclear $ 1s $ state, is relatively weaker in the light nuclear region. Since the current study employs a spherical symmetry approximation, the calculated separation energies of the $^{13}_{\Xi^{-}}$B hyperon $ 1p $ state obtained from the selected effective interaction are mostly unbound. Therefore, a detailed discussion on another potentially significant constraint, namely the separation energy of the $ 1p $ state hyperon in $^{13}_{\Xi^{-}}$B, is not conducted. Previous research has achieved a consistent description of the theoretical and experimental separation energies of the $^{13}_{\Xi^{-}}$B hyperon $ 1p $ state by considering deformation effects, assuming a deformed core of $^{12}$C in $^{13}_{\Xi^{-}}$B \cite{Guo2021PRC104.L061307}. However, there remains some model dependence on whether $^{13}_{\Xi^{-}}$B exhibits deformation effects in theory \cite{Chen2022CPC46.064109}. Thus, besides accounting for deformation effects, achieving a self-consistent description of the separation energy of the $^{13}_{\Xi^{-}}$B hyperon $ 1p $ state with experimental data may be influenced by additional aspects.

\begin{acknowledgements}
This work was partly supported by the Fundamental Research Funds for the Central Universities, Lanzhou University (lzujbky-2022-sp02, lzujbky-2023-stlt01), the National Natural Science Foundation of China (11875152, U2032141), the Strategic Priority Research Program of Chinese Academy of Sciences (XDB34000000), the Natural Science Foundation of Henan Province (242300421156).
\end{acknowledgements}

\end{document}